\documentclass[journal]{IEEEtran}
\IEEEoverridecommandlockouts
\makeatletter
\def\ps@headings{%
\def\@oddhead{\mbox{}\scriptsize\rightmark \hfil \thepage}%
\def\@evenhead{\scriptsize\thepage \hfil \leftmark\mbox{}}%
\def\@oddfoot{}%
\def\@evenfoot{}}
\makeatother
\usepackage{subfigure}
\usepackage{colortbl}
\usepackage{bm}

\usepackage{graphicx}  
\usepackage{url}       

\usepackage{amsmath}   
\usepackage{cite}

\usepackage{amsfonts,amssymb}

\usepackage{stfloats}

\usepackage{cases}
\usepackage{algorithm}
\usepackage{multirow}
\usepackage{algorithmic}
\usepackage{stfloats}
\usepackage{epstopdf}

\usepackage{xcolor}
\usepackage{xpatch}
\makeatletter
\def\changeBibColor#1{%
	\in@{#1}{}
	\ifin@\color{blue}\else\normalcolor\fi
}
\xpatchcmd\@bibitem
{\item}
{\changeBibColor{#1}\item}
{}{\fail}
\xpatchcmd\@lbibitem
{\item}
{\changeBibColor{#2}\item}
{}{\fail}

\newcommand{\Rmnum}[1]{\expandafter\@slowromancap\romannumeral #1@}

\makeatother

\newcommand{\ls}[1]
    {\dimen0=\fontdimen6\the\font
     \lineskip=#1\dimen0
     \advance\lineskip.5\fontdimen5\the\font
     \advance\lineskip-\dimen0
     \lineskiplimit=.9\lineskip
     \baselineskip=\lineskip
     \advance\baselineskip\dimen0
     \normallineskip\lineskip
     \normallineskiplimit\lineskiplimit
     \normalbaselineskip\baselineskip
     \ignorespaces
    }

\begin{document}

\title{Location-Driven Beamforming for RIS-Assisted Near-Field Communications}
\vspace{10pt}

\author{
\IEEEauthorblockN{Xiao Zheng, Wenchi Cheng,  Jingqing Wang, and Wei Zhang}

\vspace{-20pt}

\thanks{
This work was supported in part by the National Natural Science Foundation of China (No.62341132), the National Key Research and Development Program of China under Grant 2021YFC3002102, and the Key R\&D Plan of Shaanxi Province under Grant 2022ZDLGY05-09. {\it(Corresponding author: Wenchi Cheng.)}

Xiao Zheng, Wenchi Cheng, and Jingqing Wang are with Xidian University, Xi'an, 710071, China (e-mails: zheng\_xiao@stu.xidian.edu.cn; wccheng@xidian.edu.cn; jqwangxd@xidian.edu.cn).

Wei Zhang is with the School of Electrical Engineering and Telecommunications, the University of New South Wales, Sydney, Australia (e-mail: w.zhang@unsw.edu.au).
}
}

\maketitle

\begin{abstract}
Future wireless communications are promising to support ubiquitous connections and high data rates with cost-effective devices. Benefiting from the energy-efficient elements with low cost, reconfigurable intelligent surface (RIS) emerges as a potential solution to fulfill such demands, which has the capability to flexibly manipulate the wireless signals with a tunable phase. Recently, as the operational frequency ascends to the sub-terahertz (THz) bands or higher bands for wireless communications in six-generation (6G), it becomes imperative to consider the near-field propagation in RIS-assisted communications. The challenging acquisition of channel parameters is an inherent issue for near-field RIS-assisted communications, where the complex design is essential to acquire the informative near-field channel embedded with both the angle and distance information. Hence, in this paper we systematically investigate the potential of exploiting location information for near-field RIS-assisted communications. Firstly, we present the progresses in the near-field RIS-assisted communications, which are compatible with existing wireless communications and show the potential to achieve the fine-grained localization accuracy to support location-driven scheme. Then, the Fresnel zone based model is introduced, with which the location-driven beamforming scheme and corresponding frame structure are developed. Also, we elaborate on four unique advantages for leveraging location information in RIS-assisted communications, followed by numerical simulations. Finally, several key challenges and corresponding potential solutions are pointed out.

\end{abstract}

\begin{IEEEkeywords} RIS, NF-MIMO, location-driven beamforming, Fresnel zone. 
\end{IEEEkeywords}

\section{Introduction}
\IEEEPARstart{{T}}{{he}} {urgent demands for six-generation (6G) wireless networks are driven by the applications for immense throughput, ubiquitous transmissions, and massive connections for various terminals.} To meet such demands, it is essential to utilize higher frequency bands, such as the millimeter-wave (mmWave) and sub-terahertz (THz) bands, which can provide thousands of times more available bandwidth compared to the available bands in previous low-frequency wireless systems. Nevertheless, the extremely severe propagation attenuation in high-frequency bands inevitably degrades the performance by limiting coverage range and shortening transmission distance\cite{RIS11}. Massive multiple input multiple output (MIMO) is a remarkable approach to counteract signal attenuation. However, it imposes high power consumption and hardware scalability issues, thereby contradicting the objective of saving energy and resources.

{To overcome the deficiency of massive MIMO, reconfigurable intelligent surface (RIS), which can intelligently manipulate the wireless propagation environment and construct the virtual line-of-sight (LoS) propagation path, has been deemed as a promising approach to mitigate the large path loss.} The RIS typically constitutes a planar array comprising numerous passive reflective elements. Such elements consist of a phase delay unit, such as positive intrinsic negative (PIN) diodes and a patch to radiate energy. Due to its energy-efficient and cost-effective features, it can capture more incident energy by expanding its aperture without entailing large expenses and re-radiating the energy in the shape of a beam toward the desired receiver to improve the energy efficiency\cite{RIS33}. Thus, deploying the RIS in high-frequency bands is envisioned as a key technology to satisfy the future ultra-high speed services for 6G wireless communications.


With the increase of RIS size and frequency band, the transmitter and receiver generally fall within the radiating near-field region of RIS. {The near-field distance of the RIS-UE link under a MISO setup (i.e., single antenna UE and RIS array), determined by Fraunhofer distance\cite{Primer}, is hundreds of meters when the RIS area in square shape is 1 $\rm m^2$ within 30 GHz, nearly covering a typical 5G cell, let alone higher frequency bands. Especially, the near-field range in RIS-assisted MIMO systems is further expanded by jointly considering the BS's aperture and UE's aperture as well as the distances of BS-RIS and RIS-UE links\cite{Near-Field-MIMO}. Hence, near-field propagation is more likely to occur in RIS-assisted MIMO communications.} Accordingly, the spherical wave propagation brings benefits to improve the achievable rate. Firstly, the near-field channel shows natural rank-sufficient property to support simultaneous transmissions of multiple data streams, even in the LoS-dominant environment\cite{Near-Field-MIMO}. By contrast, to achieve near-optimal spatial multiplexing gain, the subarray spacing should be properly adjusted in traditional LoS-MIMO environment\cite{LOS-MIMO}. Secondly, it raises the potential to focus the energy in a specific direction and at a particular distance, referred to as beamfocusing\cite{Near-Field-MIMO}. The beamfocusing enables the improvement of energy efficiency and spatial multiplexing.

Apart from such benefits, the near-field channel also brings challenges. Therein, channel estimation (CE) is essential to exploit these superiorities. However, within the near-field region, the wave propagation is more complex and involves both angle and distance information, making it challenging to establish a clear linear relationship like that of the far-field channel. {The phase distributions of the received signals exhibit nonlinear variations among adjacent antennas and cannot be simply characterized by linear array gains based on angle-of-arrival (AOA), and the path loss coefficients for different antennas differ due to discrepancies in propagation distances.} This variability necessitates the additional parameters to accurately characterize the informative near-field channel. Under such circumstances, the low-overhead angular-domain representation based CE suited for the linear far-field channel is not valid. {To solve this issue, compressed CE schemes are developed to reduce the pilot overhead by exploiting the parametric representation of RIS-assisted near-field channel\cite{Parametric_near_field}. Therein, the channel is represented by the main propagation paths formed by several parameters (e.g., angles, distances), where the pilot length is reduced to proportional to the number of main scatterers rather than that of RIS elements. Also, decomposing the cascaded channel to derive desired separate channel parameters can be achieved by considering the geometry in the deployment phase and exploiting the quasi-static property of the array response of BS-RIS link. Such approaches essentially excavate the channel's peculiarity in terms of sparsity and geometry to reduce the overhead. However, in multi-RIS-assisted scenarios, it still necessitates complex multi-stage design to separately estimate the individual RIS-associated channel. 
}



{Nevertheless, with multiple large RISs, the virtual LoS components constructed by RISs become dominant in the overall propagation paths in future RIS-assisted communications. The geometry deployment endows a high correlation between the channel and location information. Under such circumstances, as the locations are shareable among the beamforming designs of multiple RISs, rather than necessitating separate RIS-associated channel states under CE-based schemes, they exhibit preferable flexibility over CSI and can reduce the overhead of priori information. Meanwhile, following recent advancements in localization\cite{Spherical_Localization,RIS-localization}, ultra accuracy can be achieved to further support the demand for location information.} Motivated by this, in this paper we conceive a new location-driven paradigm by capturing the property of near-field with respect to location information to address the challenges associated with CSI acquisition for future 6G RIS-assisted near-field wireless communications. {In particular, we firstly detail three recent progresses in localization techniques, which are compatible with existing wireless communications. Then, the Fresnel zone based model is introduced. Therein, we depict an example of its phase distribution on the RIS in both near-field and far-field cases. {Also, according to the Fresnel zones-based geometry, we propose the near-field location-driven beamforming scheme and design the corresponding frame structure.} Additionally, we elaborate on four unique advantages of location-driven beamforming, followed by numerical results to show its superiority in complexity and robustness. Finally, we identify several open challenges in RIS-assisted location-driven beamforming that necessitate further in-depth investigation, along with proposing potential solutions for these challenges. }
\begin{figure*}[t]
    \centerline{\includegraphics[width=14cm]{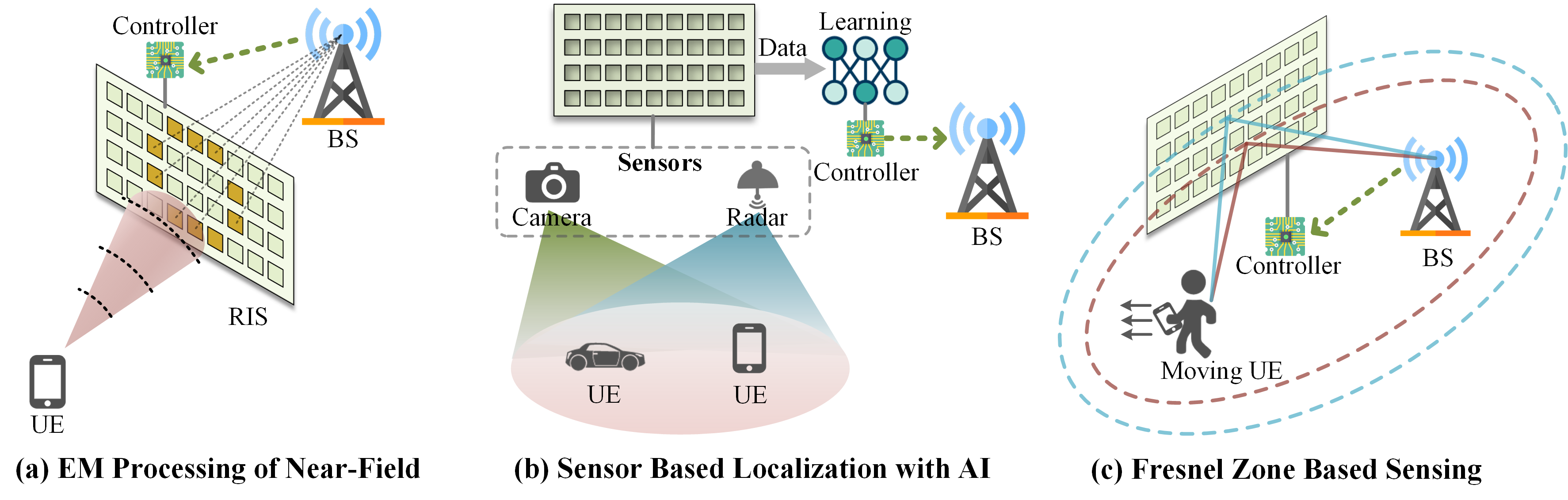}}
        \caption{Three RIS-assisted localization schemes without CE.}
        \label{localization}
        \vspace{-10pt}
\end{figure*}

{
\section{Emerging Localization Advances}
For location-driven scheme, the performance highly depends on the localization accuracy in both distance and angle. To start with, as shown in Fig.~\ref{localization}, we detail three recent progresses in localization techniques, which are compatible with existing wireless communications as these schemes mitigate the complex design of RIS, showing the potential to achieve fine-grained localization accuracy to support location-driven scheme.

\textbf{EM Processing of Near-Field: }
Within the near-field region, the wave's propagation is embedded with meaningful curvature information, offering the extra ability to acquire the exact location by processing the incoming wave. Extracting the location from the curvature information necessitates an indicator, reflecting the variation versus the wave's curvature. The authors in \cite{Spherical_Localization} proposed a general model that accounts for the electromagnetic (EM) and signal levels processing of the wavefront curvature through an EM lens. Such a lens conveys a unique relation between the incident and the output angles of the impinging and refracted waves. By adopting it to collimate the beams in precise directions, it is possible to spatially discriminate signals in the analog domain. The positioning error at about 1-2 cm can be achieved with a distance of 10 m. The bidirectional RIS is an alternative technique for the reconfigurable lens due to its low-cost and programmable control features, which can be deployed near the BS to facilitate the sensing of the UE's location in near-field RIS-assisted communications. 

\textbf{Sensor Based Localization with AI: }
Sensors (e.g., radars, lidars, and cameras) based localization is an effective technique of mobile systems. Data collected from multiple sensors provide valuable information about the desired target stemming from the same environment. AI-based techniques can effectively extract the features of the acquired data through nonlinear neural networks and map them to real locations by exploiting the interplay between sensed data and UE's location. The authors in \cite{Vision-Aided} proposed a computer vision (CV)-aided framework, where a camera attached to the BS captures the image and the deep learning-based object detector identifies the 3D location of the mobile UE. The sensors can also be integrated on RIS to directly estimate location using only the phase information of signals received from the UE with over 99 percent accuracy in determining the receiver position\cite{Sensor-based}, which does not require the exchange of CSI between UE and the Tx through pilot signals, thus increasing the proportion of information transmission in one frame. Essentially, AI-based techniques act as a black box with feature extraction and data fusion functions\cite{RIS22}. Hence, the priori information such as the coarse location based on global navigation satellite system (GNSS) may be valuable to shrink the search space to promote the localization efficiency.

\textbf{Fresnel Zone Based Sensing: }
Fresnel zones refer to the series of concentric ellipsoids with two foci corresponding to the transmitter and receiver antennas, where radio waves traveling through the odd and even Fresnel zones are in-phase and out-phase, respectively. Successive Fresnel zones alternately provide destructive and constructive interference to the received signal strength at the UE. The Fresnel zone model reveals the fluctuation of received signal caused by the motion of the target. The physical parameters of the Fresnel ellipsoids depend on the distance of Tx-Rx link and the operational frequency. Hence, this model is typically integrated with multicarrier system (e.g., WiFi) to select favorable distributions of the Fresnel ellipsoids to improve the detection performance. Therein, by deploying a pair of WiFi transceivers at the fixed location, the UE's motion induces the variation of reflected wave, from which the location and moving direction can be extracted from the amplitude and frequency of the fluctuation\cite{Fresnel2}. Deploying RIS in such a system has the potential to expand the detection range and alleviate the requirement for a pair of transceivers, as the RIS can act as a virtual Tx by flexibly reflecting the signals.

In summary, the RISs introduce the potential of achieving high-precision localization and sensing capabilities, which are standardized by European Telecommunications Standards Institute (ETSI)\cite{RIS_standard}. Meanwhile, from 3GPP Release 16 to Release 18, the performance metrics used for evaluating performance of new radio (NR) positioning including horizontal accuracy, vertical accuracy, and latency level progressively increase. These standardizations significantly facilitate the implementation of location-driven beamforming. }

\section{Location-Driven Framework}
{The interaction between the Fresnel zone based model and environment not only enable the capability to sense the UE's location, but also can be exploited to design the beamforming without CE. In this section, we first detail the property Fresnel's ellipsoid in near-field RIS-assisted communications, with which the location-driven beamforming scheme is proposed, followed by the corresponding frame structure. }

\subsection{Fresnel Zone Based Model}
In this section, we unite the Fresnel zone based model with the RIS to show their interplay in near-field RIS-assisted communications.

\begin{figure}[t]
    \centerline{\includegraphics[width=7.5cm]{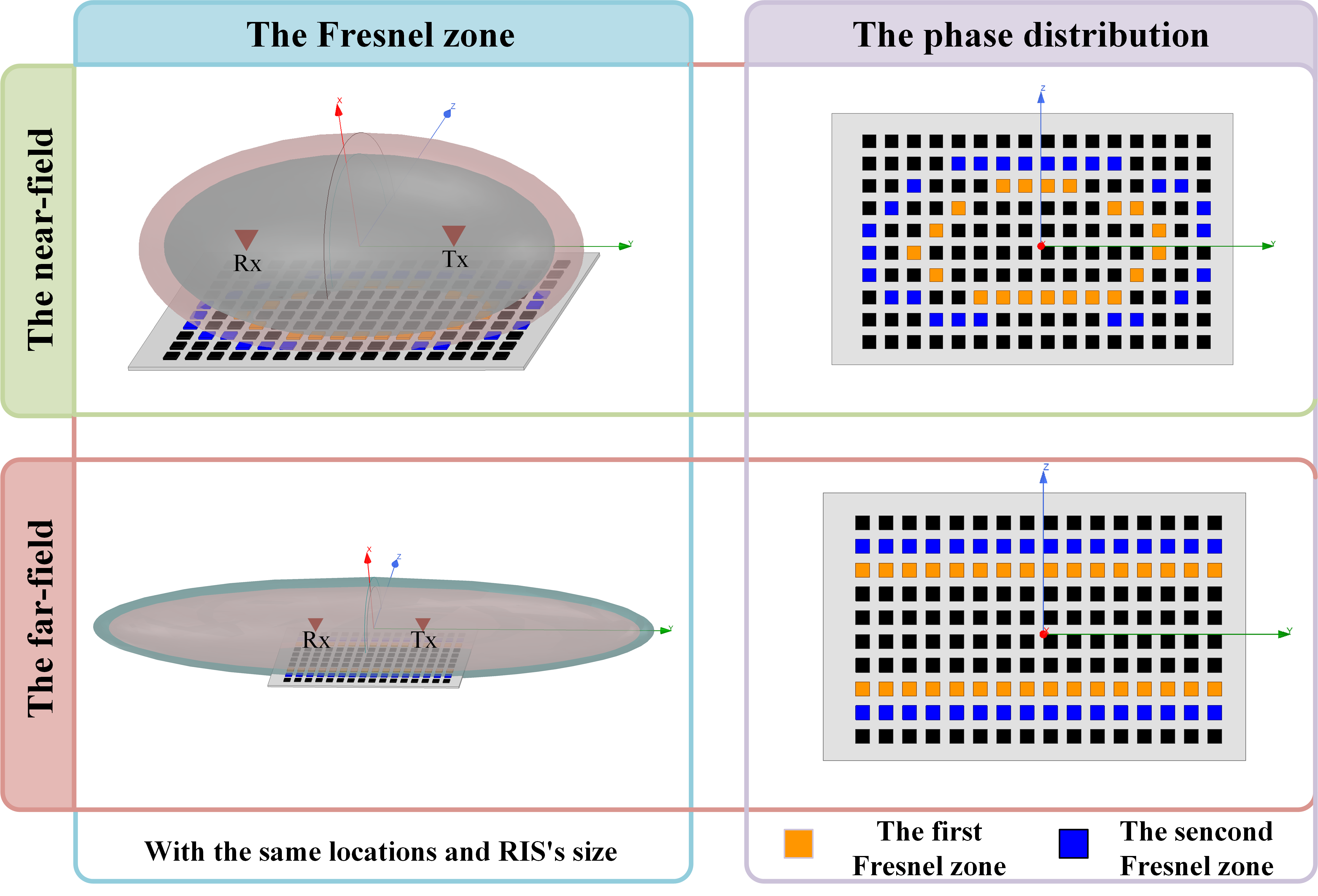}}
        \caption{The schematic diagrams of Fresnel zones and phase distributions on the RIS in the near-field and the far-field.}
        \label{Fresnel}
\end{figure}

Within the near-field region, the equiphase surfaces of the transmit signals, defined as the set of points at which a specified component of one of the vectors in a traveling electromagnetic wave has the same phase at the same time, can be characterized by the Fresnel zone. Based on the diffraction theory, the Fresnel zone describes the interference phenomenon between the reflected signal and the direct signal. It is composed of multiple ellipsoids, where the reflected signals from these ellipsoids result in constructive or destructive interferences successively. The schematic diagrams of Fresnel zone in both near-field and far-field are depicted in Fig.~\ref{Fresnel}, along with the phase distributions on the RIS. As shown in Fig.~\ref{Fresnel}, with the same scale and locations, the Fresnel zone in the far-field is narrower and longer than that of the near-field case. As the frequency increases, the semi-major axis of the Fresnel ellipsoid (i.e., Fresnel radius) increases, leading to different phase distributions on the RIS. For the far-field case, the phase distributions of the first and the second Fresnel zones, represented by elements with yellow and blue, respectively, form nearly parallel lines. In contrast, for the near-field case, the phase distributions take the shape of an ellipse, whose curvature contains the information about the wave's direction. The Fresnel zone based model implies an interaction between the locations and channel parameters\cite{Fresnel2}, which provides a clear elucidation for the RIS's behavior in the near-field.

\subsection{Location-Driven Beamforming}
\begin{figure}[t]
    \centerline{\includegraphics[width=7.5cm]{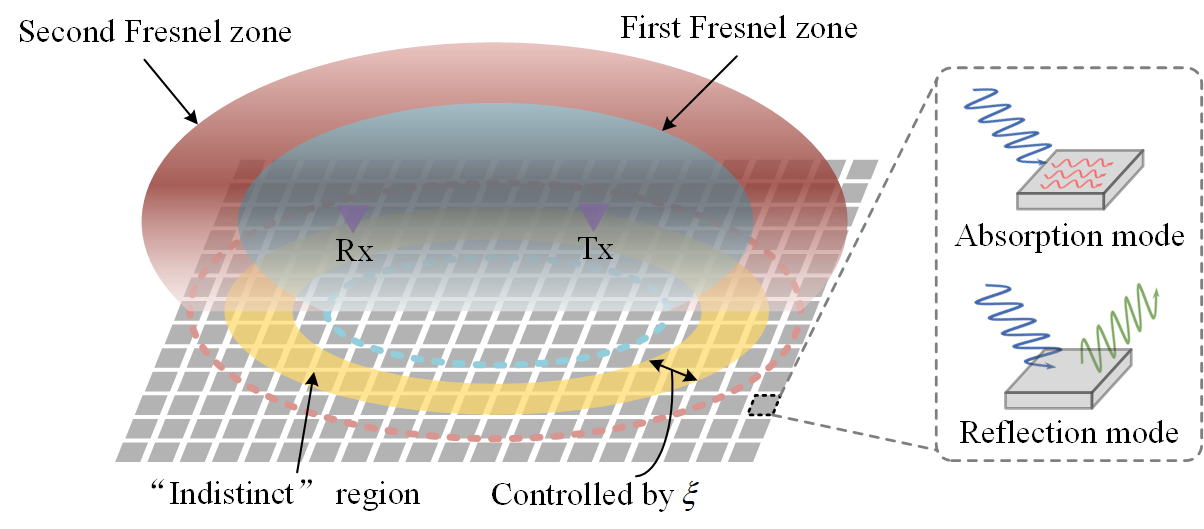}}
        \caption{{The geometric relations formed by Fresnel zone based model.}}
        \label{Fresnel_zone}
        \vspace{-10pt}
\end{figure}

In the near-field, the equiphase surfaces impinging on RIS, generated by the interaction between Fresnel zone ellipsoids and the RIS plane, manifest as a collection of curves, as shown in Fig.~\ref{Fresnel_zone}. {Motivated by this, we exploited the curvatures of the Fresnel ellipsoids to manipulate the phase shift of 1-bit RIS based on location information. The curvatures of these curves contain abundant propagation information, which can be described by ellipsoid equations. The distributions of these ellipsoids depend on the distance of Tx-Rx link and operational frequency. Based on the known locations, simultaneous equations of the ellipsoids and the RIS plane are established and addressed by coordinate transformation. The solutions of the simultaneous equation are the intersecting region between Fresnel ellipsoids and RIS plane. After solving the formed simultaneous equations with respect to various indexes of Fresnel zones, the phase distribution over such region can be derived and exploited to design the configuration. }
\begin{figure*}[t]
    \centerline{\includegraphics[width=14cm]{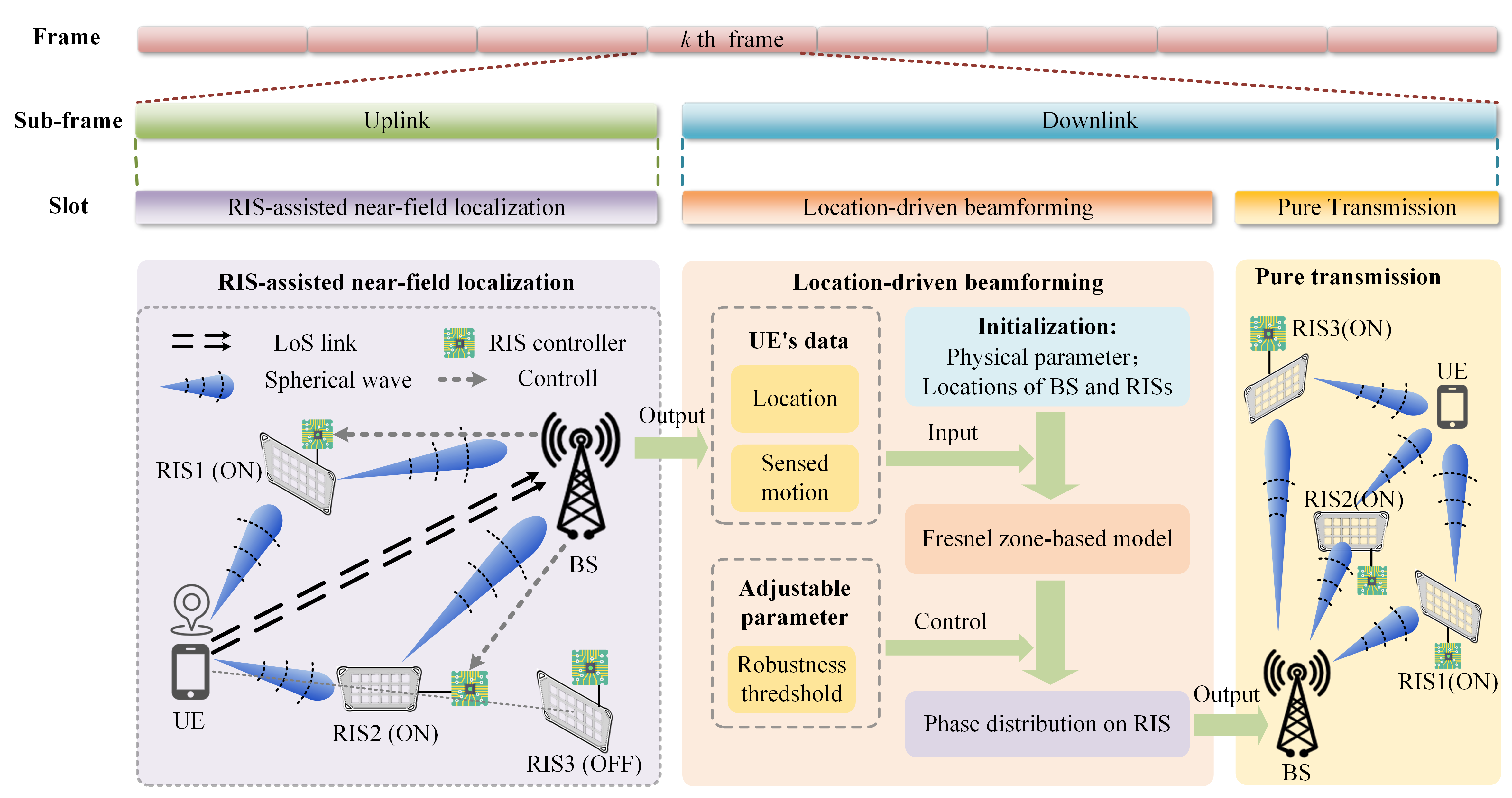}}
        \caption{The time-division frame structure in location-driven beamforming.}
        \label{frame}
        \vspace{-10pt}
\end{figure*}

{The above solutions are effective in the ideal scenario in the absence of location error. Nevertheless, the location uncertainty always exists in practical scenarios. This means that the derived phase distributions are erroneous, which will greatly degrade the performance. To alleviate this deficiency, we propose the two-step position-aided On/Off state judgement (TPOSJ) scheme\cite{Xiao_zheng2}. In specific, inspired by the property of Fresnel zone, these elements located at the center between two adjacent Fresnel zones are denoted as ``indistinct'', as shown in Fig.~\ref{Fresnel_zone}. These elements are sensitive to location deviation, causing wrong beamforming configurations. Considering the worst case, the erroneous configurations lead to that reflected signals from these elements being all out-phase with received signal, which greatly degrades the received power. Hence, we design the two-mode element with joint absorption and reflection modes to alleviate the impact of wrong configurations.} The element in reflection mode reflects the incident wave with a controllable phase delay, while the element in absorption mode fully absorbs the energy of incident wave without reflection. Consequently, these elements in the ``indistinct'' region are configured in absorption mode. {Such absorption mode has been referred in ETSI standards\cite{RIS_standard}, but its use cases are not yet explored. In the TPOSJ scheme, it is used to improve the robustness against location error.} {Furthermore, to derive a specific tradeoff between robustness and achievable rate according to the performance constraints, we design the adjustable threshold, denoted by $\xi$, to determine the width of the ``indistinct'' region. With smaller $\xi$, there are more elements configured in absorption mode and vice versa. Its value varies from 0 to $\lambda/2$, where all elements are configured in reflection mode and absorption mode with $\xi=\lambda/2$ and $\xi=0$, respectively. Such threshold can be determined according to the degrees of location errors.}

Furthermore, with the TPOSJ scheme, we propose the location-driven frame structure in Fig.~\ref{frame}. {As shown in Fig.~\ref{frame}, with the time-division-duplexing (TDD) protocol, each frame is divided into uplink and downlink. {In the uplink of the $k$th frame, the UE transmits its signals by using the configuration same as that of the $\left(k-1\right) $th downlink based on channel reciprocity.} Three previously mentioned RIS-assisted localization techniques as shown in Fig.~\ref{localization} can be used in this phase. It is noteworthy that unless using the localization scheme in Fig.~\ref{localization} (a), the pilot signal is not indispensable in uplink stage. Then, during the downlink, the BS extracts the location information from the received signals and input the sensed locations into TPOSJ scheme. }Based on the physical parameters and locations, the location-driven beamforming problem is formulated and solved according to the proposed TPOSJ scheme, where the threshold $\xi$ can be dynamically controlled according to the performance constraints. Finally, the configuration of RIS is conveyed through control signalling from BS, followed by pure data transmission. Therein, the network-controlled RIS only needs to receive the control information from the network\cite{RIS_standard} in the absence of pilot transmission. {The same configuration is also used for the transmission in the uplink of the $\left(k+1\right) $th frame.}

\section{Advantages of Location-Driven Beamforming}
{In this section, we elaborate on four significant advantages of exploiting location information for RIS-assisted near-field communications. } 

\textbf{High precision localization: }
The deployment of RIS introduces tremendous advantages for traditional radio localization, especially in the near-field region. In specific, under the rigid requirement for synchronization among BSs, RIS-enabled localization shows natural synchronization\cite{RIS-localization}. With the synchronized cost-effective RISs, the required number of BSs can be drastically reduced, even in NLoS condition. In contrast, in conventional localization without implementing the RIS, four BSs are needed to measure the clock bias by time-difference-of-arrival (TDoA) and solve the 3D positioning and 1D clock bias estimation problem\cite{RIS-localization}. {Also, the latest advances, as shown in Fig.~\ref{localization}, has the potential to achieve fine-grained localization accuracy. For instance, the positioning error at about 1-2 cm can be achieved with a distance of 10 m in the near-field region\cite{Spherical_Localization} while the centimeter and even millimeter scale motion can be sensed\cite{Fresnel2}.
}

\textbf{Mobility support: }
The UE's mobility can be naturally captured in location-driven frame by trajectory prediction techniques and feedback mechanisms. Especially for future Internet of Everything (IoE)-based smart applications, the motion and velocity are always sensed and monitored. Additionally, in RIS-assisted near-field communications, the RIS enriches the scattering environment and helps to improve the precision. The detection of subtle movements (e.g., respiration detection and gesture recognition) can be achieved by leveraging only the Fresnel zone based model by extracting the fluctuation of received signal within the near-field region\cite{Fresnel2}. {It is feasible to unite the sensed parameters with prediction techniques (e.g., Kalman prediction, linear prediction) to predict UE's next location. Such predicted parameters can be exploited in location-driven scheme to overcome the deficiency of acquirement of outdated locations. 

{\textbf{Flexible configuration: }
The pilot overhead of estimation depends on the selective target (i.e., CSI or location) and its corresponding unknown parameters. Specifically, for CE-based schemes, although parametric representations of the channel can effectively reduce the pilot length, the estimation overhead is still proportional to the number of RISs because the targets are separate RIS-associated channels. In contrast, the target in the localization stage is the UE's location, which means that the pilot overhead may not increase with variations in the number of RISs. For instance, the RISs with the similar direction relative to the UE cannot significantly improve the localization performance (e.g, RIS2 and RIS3 as shown in Fig.~\ref{frame}), and thus only enabling one of them helps to reduce the pilot overhead. Also, one RIS lens is sufficient to estimate the location at a desirable accuracy in the near-field\cite{Spherical_Localization}. Hence, in practice, several RISs with preferable condition can be enabled while switching off the others to reduce the pilot overhead. Then, in the communication stage, all RISs can be enabled to enhance the transmission rate.}


{\textbf{Favorable compatibility: }
Driven by a wide range of emerging applications such as automated vehicles and robots, future wireless networks require high-resolution environmental awareness. Such capability aligns precisely with the needs of the location-driven framework. The advanced sensing and localization functions that determine the UE's location can be effectively applied to the uplink frame of the proposed framework. More importantly, under the same theoretical framework, uniting the Fresnel zone based sensing\cite{Fresnel2} with the proposed location-driven scheme gives an attractive example for integrated sensing and communication (ISAC). Therein, the Fresnel zone based sensing only depends on the fluctuation of received signal, which can be united with localization techniques to further facilitate accuracy. Accordingly, by inputting the sensed UE's location into the proposed location-driven scheme, the communication performance can be enhanced.
}

\section{Numerical Results}
To further demonstrate the advantages of location-driven beamforming, we present numerical results of comparing the computational complexity and spectrum efficiency of the proposed location-driven beamforming. {Unless specified otherwise, the RIS is composed of $80\times80$ elements with half a wavelength size operating in 28 GHz, the locations of transmitter and receiver are $(0, 12, 0)$ and $(5, 0, 0)$ in Cartesian system, respectively. The transmit power and noise power are set as 30 dBm and -90 dBm, respectively.}

\begin{figure}[t]
    \centerline{\includegraphics[width=8cm]{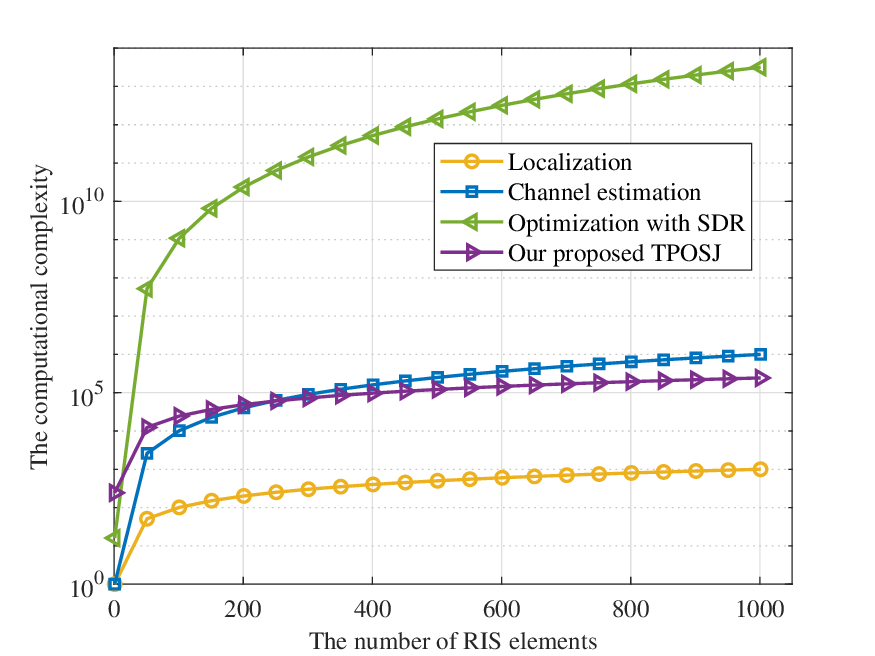}}
        \caption{Complexity comparison between location-driven beamforming and CSI-driven beamforming.}
        \label{complexity}
        \vspace{-15pt}
\end{figure}

Figure \ref{complexity} plots the computational complexities of localization method, cascaded CE with multiuser joint channel estimation (MJCE), the passive beamforming with the semi-definite relaxation (SDR)\cite{CSI-free}, and our proposed TPOSJ scheme versus the number of RIS elements. As shown in Fig.~\ref{complexity}, there are two significant gaps between the two pairs. {The gap between CE and localization arises because the estimated target in location-driven beamforming is the UE's location, rather than RIS-associated links. Hence, the RISs with preferable conditions can be enabled so that the complexity of localization can be reduced by switching off the profitless RISs. Also, the computational complexity gap between the optimization with SDR and our proposed TPOSJ is large. This is because, by integrating the Fresnel zone with location information, the beamforming configuration does not require an optimization format, thus greatly reducing the complexity\cite{Xiao_zheng2}.} 

\begin{figure}[t]
    \centerline{\includegraphics[width=8cm]{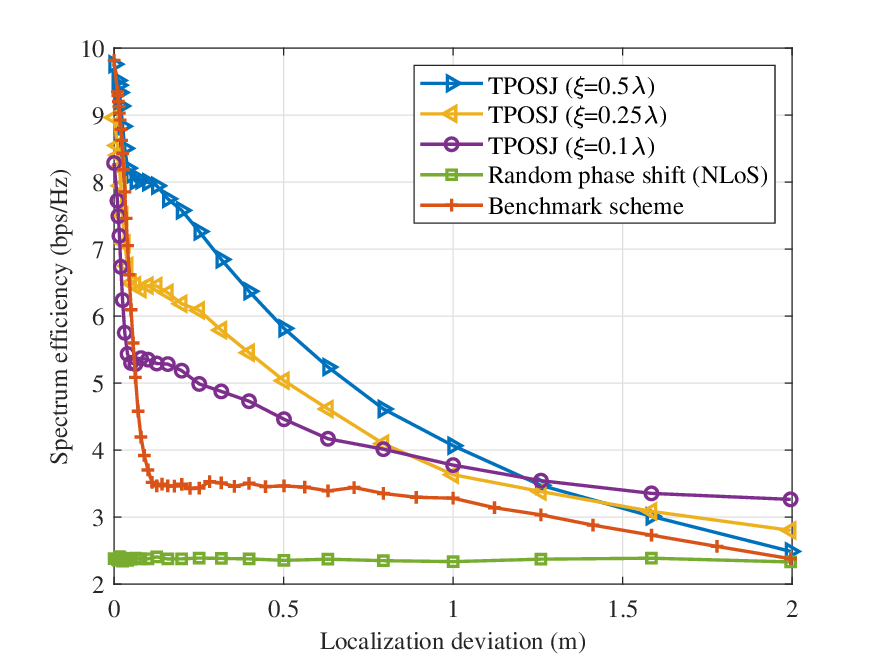}}
    \caption{{The spectrum efficiency versus location deviation in NLoS environment.}}
    \label{xi_tu}
    \vspace{-15pt}
\end{figure}

{Figure~\ref{xi_tu} shows the spectrum efficiency versus location deviation including the TPOSJ scheme with $\xi$ being $0.1\lambda$, $0.25\lambda$, and $0.5\lambda$, respectively, and the benchmark scheme.} The location is directly exploited in the benchmark scheme by using the free-space transmission equation without absorption mode element. As shown in Fig.~\ref{xi_tu}, the spectrum efficiencies of these two schemes drop from the same starting point when $\xi=0.5\lambda$. With the increase of location error, the TPOSJ scheme first rapidly degrades in low error regime, and later slowly degrades until approaching the performance of random phase shift, which contains a relatively stable interval. In contrast, the spectrum efficiency of the benchmark scheme decreases more seriously as the location error increases, which also finally approaches the performance of random phase shift. Also, the spectrum efficiency in the absence of location error increases as the robustness threshold increases. In low location error region, the performance with higher value of $\xi$ surpasses that of with lower value. That is, increasing $\xi$ can improve the performance in low location error region. Nevertheless, as $\xi$ decreases, the rate of decline slows down and the stable interval extends. This phenomenon occurs because the reflected signals from elements in reflection mode exhibit a robust constructive superposition, which is less influenced by localization deviations. Conversely, in high location error regime, the performance with lower value of $\xi$ surpasses that of with higher value. Decreasing $\xi$ can improve the performance in high location error region. Thus, the absorption mode can be used to improve the robustness for the scenario where the localization accuracy is coarse or Tx is in motion with mobility. The tradeoff between the maximum achievable rate and robustness to location error can be dynamically adjusted based on the degree of location error.

\section{Challenges for Location-Driven Beamforming}
Despite the above advantages, there are some challenges regarding applying the location-driven beamforming in RIS-assisted communications. In this section, we discuss these challenges and provide some potential solutions.

\textbf{Tradeoff between localization precision and performance: }
{The high location uncertainty will undoubtedly degrade the overall performance of location-driven scheme. Particularly, both precise angle and distance information is needed in the near-field. The demands for dual information of angle and distance increases the risks of suffering from ``barrel law''. Therein, the performance is limited by the information (i.e., angle and distance) with larger error. Once one of the angle and distance information is ambiguous, the energy will be focused at an undesirable location, thus degrading the energy efficiency and causing signal leakage. Higher location precision necessitates more resources to facilitate the localization, reducing the transmission efficiency. Hence, elegant tradeoff between localization precision and performance is highly desired. }In our work, we find that the affected degree of performance versus different coordinates components of location errors are discrepant. Thus, localization with coarse and fine precision corresponding to different coordinates components has the potential to solve this issue. 

{\textbf{Robust design: }
The performance of location-driven beamforming suffers from UE's unavoidable location error. The blind area exists, such as the complex indoor scenarios, where diffraction becomes an important phenomenon in the absence of LOS paths due to high wall/floor attenuation. Under such harsh environments, a certain localization accuracy cannot be attained. {Additionally, the UE's unpredictable mobility also results in the acquisition of outdated locations in location-driven scheme, bringing the performance degradation analogous to that caused by location error.} Hence, it is imperative to design the robust beamforming scheme to prevent the transmission outage considering the worst case. Such problem is initially considered by authors in \cite{CSI-free}, where they proposed a relaxed alternating optimization process to solve the worst-case robust beamforming optimization problem according to the location error region. Also, in our proposed TPOSJ scheme, we design the element with absorption mode to combat the location error. Efficient robust beamforming schemes still need further explorations.

\textbf{Mapping from location to channel: }
{With the increase in frequency and the deployment of RISs, the sparse propagation environment highly relates to the distributions of RISs and several main scatterers.} Under this condition, the location information effectively correlates with the channel's state so that it can accurately reproduce the entire channel. {Nevertheless, the mapping from location to channel is not valid in the complex electromagnetic environment (e.g., indoor scenario). The high attenuation induced by wall seriously limits the localization accuracy and the deviation of mapping from location to channel brings the same effect analogous to location error. Such deviation can be mitigated by robust design but cannot be radically addressed. With the development of multiple large RISs in future 6G networks, especially the holographic RIS, the reconfigurable paths from RIS dominate the propagation so that the location-driven scheme can be significantly facilitated with unique advantages.}
}

\section{Conclusions}
In this paper, we systematically explored the potential of employing the location information for beamforming design to address the challenges associated with CSI acquisition for future 6G RIS-assisted near-field wireless communications. {Specifically, we firstly detailed three recent progresses in localization techniques, which are compatible with existing wireless communications. Then, the Fresnel zone based model is introduced. Therein, we depicted an example of its phase distribution on the RIS in both near-field and far-field cases. Also, according to the Fresnel zones-based geometry, we proposed the near-field location-driven beamforming scheme and design the corresponding frame structure. Additionally, we elaborated on four unique advantages of location-driven beamforming, followed by numerical results to show its superiority in complexity and robustness. Finally, we identified several open challenges in RIS-assisted location-driven beamforming, along with proposing potential solutions for these challenges. In conclusion, the investigations corresponding to location-driven beamforming are still in the preliminary stage, which necessitates further in-depth investigation to address these inherent challenges. }

\bibliography{References}
\bibliographystyle{ieeetr}
\vspace{-17pt}
\begin{IEEEbiography}[{\includegraphics[width=1in,height=1.25in,clip,keepaspectratio]{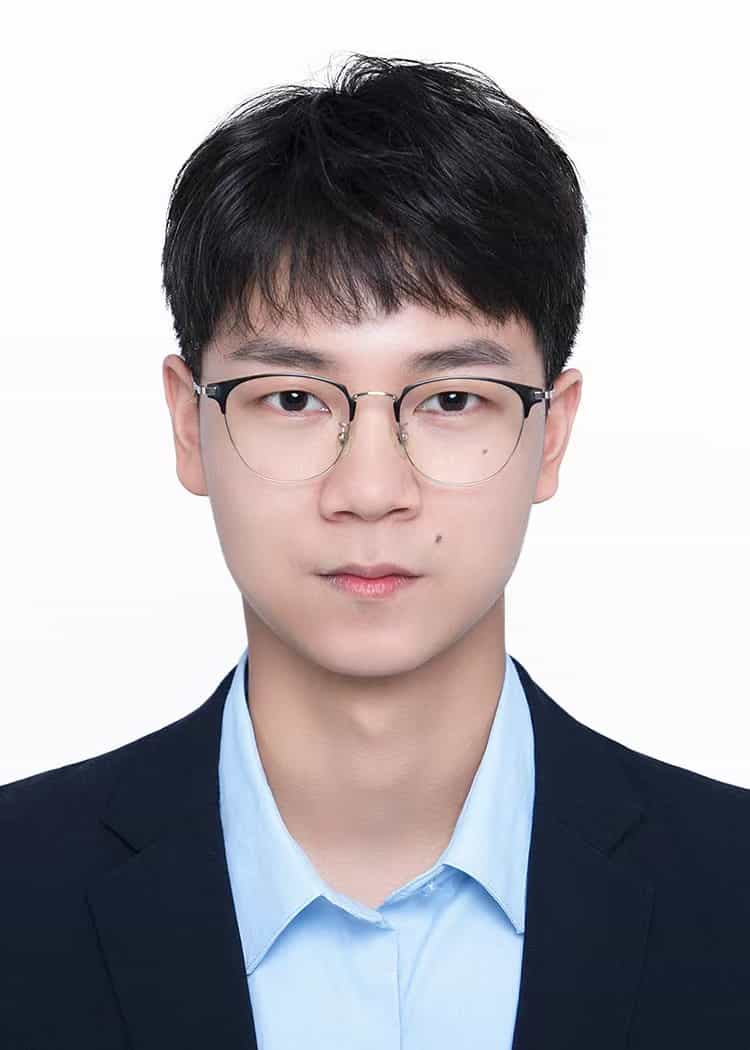}}]
{Xiao Zheng} received the B.S. degree in telecommunication engineering from Xidian University, China, in 2020. He is currently pursuing the Ph.D. degree in telecommunication engineering at Xidian University. His research interests focus on near-field communications and reconfigurable intelligent surface.
\vspace{-17pt}
\end{IEEEbiography}

\begin{IEEEbiography}[{\includegraphics[width=1in,height=1.25in,clip,keepaspectratio]{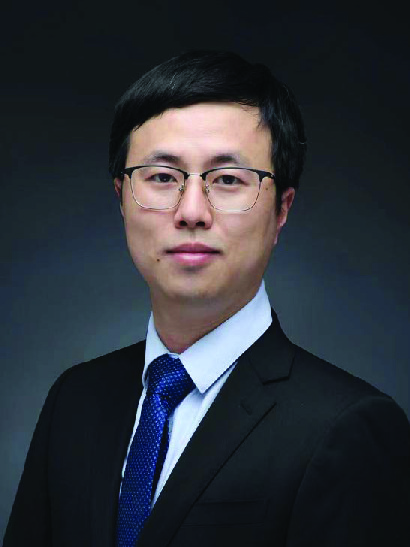}}]{Wenchi Cheng} received his B.S. and Ph.D. degrees in telecommunication engineering from Xidian University in 2008 and 2013, respectively, where he is a full professor. He was a Visiting Scholar with the Department of Electrical and Computer Engineering, Texas A\&M University, College Station, TX, USA, from 2010 to 2011. He has published more than 150 international journal and conference papers in the IEEE JSAC, IEEE magazines, and IEEE transactions, and at conferences including IEEE INFOCOM, GLOBECOM, ICC, and more. He received the IEEE ComSoc Asia-Pacific Outstanding Young Researcher Award in 2021, the URSI Young Scientist Award in 2019, the Young Elite Scientist Award of CAST, and four IEEE journal/conference best papers. His current research interests include emergency wireless communications and orbital-angular-momentum-based wireless communications. 
\end{IEEEbiography}
\vspace{-17pt}

\begin{IEEEbiography}[{\includegraphics[width=1in,height=1.25in,clip,keepaspectratio]{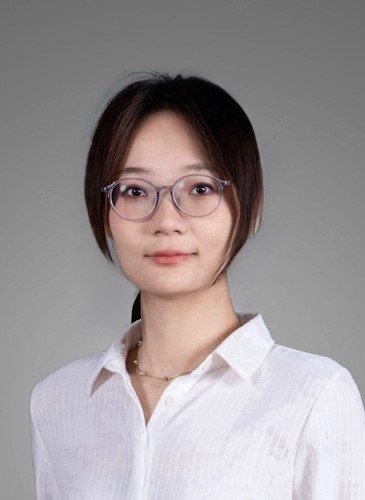}}]{Jingqing Wang} received the B.S. degree from Northwestern Polytechnical University, in Electronics and Information Engineering and the Ph.D. degree from Texas A\&M University, College Station, in Computer Engineering in 2022. She is currently a Lecturer with Xidian University. She has published more than 60 international journal and conference papers in IEEE JOURNAL ON SELECTED AREAS IN COMMUNICATIONS, IEEE magazines, IEEE TRANSACTIONS, IEEE INFOCOM, GLOBECOM, WCNC, and ICC. She won the Best Paper Award from the IEEE GLOBECOM in 2020 and 2014, respectively. Her current research interests focus on next generation mobile wireless network technologies, statistical delay and error-rate bounded QoS provisioning, 6G mURLLC, information-theoretic analyses of FBC, emerging machine learning techniques.
\end{IEEEbiography}
\vspace{-17pt}

\begin{IEEEbiography}[{\includegraphics[width=1in,height=1.25in,clip,keepaspectratio]{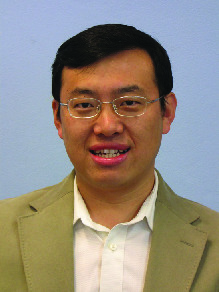}}] {Wei Zhang} received the Ph.D. degree in electronic engineering from The Chinese University of Hong Kong in 2005. He is currently a Professor with the School of Electrical Engineering and Telecommunications, University of New South Wales, Sydney, NSW, Australia. He has published more than 200 articles and holds five U.S. patents. His research interests include millimetre wave communications and massive MIMO. He is the Vice Director of the IEEE ComSoc Asia Pacific Board. He serves as an Area Editor for the IEEE TRANSACTIONS ON WIRELESS COMMUNICATIONS and the Editor-in-Chief for \emph{Journal of Communications and Information Networks}.
\end{IEEEbiography}

\end{document}